\documentclass[conference]{IEEEtran}
\IEEEoverridecommandlockouts
\usepackage{cite}
\usepackage{amsmath,amssymb,amsfonts}
\usepackage{algorithmic}
\usepackage{graphicx}
\usepackage{graphics}
\usepackage{booktabs}
\usepackage{multirow}
\usepackage{textcomp}
\usepackage{xcolor}
\usepackage{todonotes}
\usepackage{caption}
\usepackage{multirow}
\usepackage[symbol]{footmisc}
\usepackage{soul}
\renewcommand{\baselinestretch}{0.9}
\def\BibTeX{{\rm B\kern-.05em{\sc i\kern-.025em b}\kern-.08em
    T\kern-.1667em\lower.7ex\hbox{E}\kern-.125emX}}
\begin{document}

\title{Deep Learning-based Embedded Intrusion Detection System for Automotive CAN%
}

\author{\IEEEauthorblockN{Shashwat Khandelwal, Eashan Wadhwa \& Shanker Shreejith}
\IEEEauthorblockA{Department of Electronic and Electrical Engineering,
Trinity College Dublin\\
Dublin, Ireland\\
Email: \{khandels, wadhwae, shankers\}@tcd.ie}}

\maketitle

\begin{abstract}
Rising complexity of in-vehicle electronics is enabling new capabilities like autonomous driving and active safety. 
However, rising automation also increases risk of security threats which is compounded by lack of in-built security measures in legacy networks like CAN, allowing attackers to observe, tamper and modify information shared over such broadcast networks. 
Various intrusion detection approaches have been proposed to detect and tackle such threats, with machine learning models proving highly effective. 
However, deploying machine learning models will require high processing power through high-end processors or GPUs to perform them close to line rate.
In this paper, we propose a hybrid FPGA-based ECU approach that can transparently integrate IDS functionality through a dedicated off-the-shelf hardware accelerator that implements a deep-CNN intrusion detection model.
Our results show that the proposed approach provides an average accuracy of over 99\% across multiple attack datasets with 0.64\% false detection rates while consuming 94\% less energy and achieving 51.8\% reduction in per-message processing latency when compared to IDS implementations on GPUs. 
\end{abstract}


\section{Introduction \& Related Works}\label{sec:introduction}
New generation of vehicles employ over 50 distributed electronic control units (ECUs) and are more connected to the outside world than ever before to offer safety-critical, convenience and infotainment services. 
Though the internal networks and electric/electronic (E/E) architectures evolved over time, they are still implemented in a modular fashion with the ECUs exchanging sensor and control information over the internal networks. 
While E/E architectures were originally partitioned and isolated into critical and non-critical functional domains, with different networks catering to the requirements of the domain, bridged access (through the on-board diagnostic port among others) were also established due to regulatory and maintenance requirements.
The increased connectivity to monitor and control critical systems for remote diagnostics, software services and user accessibility has further reduced this isolation.
This opens new avenues for attackers to deploy both invasive and non-invasive schemes to inject malicious content on these previously siloed internal networks~\cite{nie2017free,iehira2018spoofing,cai20190}. 
Such attacks range from simple fuzzing and replay attacks to more intelligent spoofing and denial of service attacks leading to total loss of critical controls in the vehicle affecting safety of users and property. 
These attacks are feasible since most vehicular networks in use today, like Controller Area Network (CAN), offer no mechanism to authorise or validate ECUs or messages on the network, other than physical connectivity.

To counter these attacks, security approaches have been explored at the logical level, specifically for legacy networks like CAN~\cite{alshammari2018classification,yang2019tree,song2020vehicle,tariq2020cantransfer,desta2020mlids,narayanan2015using}. 
Packet inspection techniques rely on checking for abnormalities in all messages transferred on the CAN bus and to alert the ECU in case of a potential malware~\cite{ling2012algorithm,narayanan2015using,weber2018embedded}.
More focused approaches like Intrusion Detection Systems (IDSs) look for network intrusions by correlating network traffic patterns and message timings to detect onset of malicious activity. 
Early flavours of IDSs correlated previously known malicious packet signatures with active messages on the CAN bus to detect intrusions through rule-based classification; however, their success rate in identifying new and evolving attack vectors were poor and incurred incremental storage each time a new signature was identified. 

More recently, machine learning (ML) models have shown promising results owing to their generalisable nature. 
Classical ML techniques like SVMs~\cite{alshammari2018classification} \& tree based approaches~\cite{yang2019tree} have been proposed as IDSs.
Various deep learning approaches have also been proposed as IDSs. 
In~\cite{song2020vehicle}, the authors propose a reduced inception net architecture for IDS that uses deep convolutional neural networks. 
The authors show that the ML architecture can achieve over 99\% accuracy across DoS, fuzzing and spoofing attacks.
The authors used a dataset captured from actual vehicle for training and testing their model, which has been shared with the community for further research.
Since the dataset covers multiple attack modes with actual CAN messages, we use the same dataset to train and evaluate our proposed architecture. 
Other approaches include the use of long-short-term memory (LSTM)~\cite{desta2020mlids,hanselmann2020canet}, convolutional LSTMs with transfer learning~\cite{tariq2020cantransfer} \&  generative adversarial networks (GAN) to detect attack signatures. 
All the approaches use CAN IDs, payload or the entire frame as their input feature.
In~\cite{de2021efficient}, the authors use an iForest anomaly detection algorithm to detect fuzzing and spoofing (RPM \& Gear) attacks and mark the message as an error preventing its propagation to other ECUs; however this can cause multiple messages to be dropped from the bus in case of false positives or DoS attacks.

Despite the promising results, deploying them in a E/E system can be challenging. 
Software implementations of ML-models for line-rate intrusion detection will have to share compute resources with critical tasks on a standard ECU leading to isolation issues and high computational overheads. 
Complex ML-based IDSs are hence proposed as standalone ECUs or as loosely-coupled GPU accelerators for line rate detection.
However, the higher power consumption coupled with factors like weight, size and integration complexity are a strong barrier towards widespread adoption of dedicated IDS ECUs, particularly in distributed IDS deployments. 
Alternatively, an ECU architecture that closely couples a specialised hardware accelerator on the same die could be used to offload the intrusion detection task, where a hardware-efficient version of the ML-network is executed exploiting its parallelism. 
Hybrid FPGA devices like the Zynq Ultrascale+ from Xilinx allows for such an ECU architecture, with capable ARM processors and a programmable fabric closely integrated on the same die, offering custom parallelism at much lower power consumption. 
Prior research has explored the case for FPGA-based ECUs to enable compute acceleration of complex tasks in vehicular systems~\cite{cho2021fpga,shreejith2013reconfigurable}.
Hybrid FPGA-based ECUs have also been developed by automotive vendors offering improved functional consolidation and reliability~\cite{BoschMS6}.
However, integrating efficient accelerators closely coupled with software functions on the ECU requires custom designs with low-level optimisations in hardware, software and interfaces~\cite{shreejith2018smart}.

In this paper, we explore a tightly coupled IDS accelerator on the Zynq Ultrascale+ platform that implements a hardware efficient ML accelerator. 
We use a quantised deep-CNN (QdCNN)-based intrusion detection model that is executed on an off-the-shelf Xilinx's deep learning processing unit (DPU) accelerator IP attached as a slave peripheral to the ECU, allowing it to be driven and controlled through software APIs from the  ECU application. 
This approach enables the ECU to execute its normal tasks and offload the ML-IDS function to the accelerator when a new message is received from the CAN interface.
The QdCNN IDS achieves an average accuracy of 99.32\% across multiple attack vectors such as Denial of Service (DoS), Fuzzy, and spoofing (RPM and Gear) attacks, identical to or exceeding the detection accuracy achieved by state-of-the-art \text{GPU- and CPU-based} implementations. 
Our experiment shows that the tight integration reduces the per message execution latency by 51.8\% and the power consumed by 94\% compared to GPU implementation of the QdCNN network, while also bettering results from competing techniques in literature.
Furthermore, the low resource overhead of the architecture also enables performance scaling through parametric parallelism at each DPU and through multiple DPU instances on the same ECU. 
The remainder of the paper is organised as follows. Section~\ref{sec:proposedmodel} describes the proposed deep-CNN model and the implemented architecture on the Ultrascale+ device; section~\ref{sec:experiments} outlines the experiment setup and results; and we conclude the paper in section~\ref{sec:conclusion}. 

\section{System Architecture}\label{sec:proposedmodel}
\subsection{IDS ECU Architecture}
\begin{figure}[t!]
    \centering
    \includegraphics[scale = 0.61]{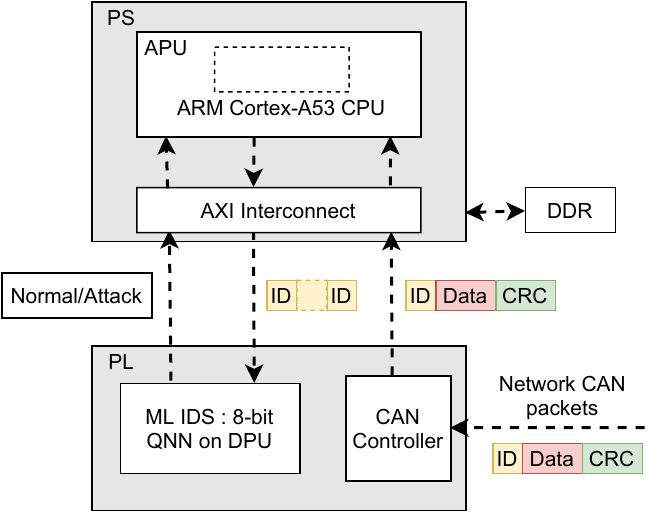}
    \caption{The proposed IDS-ECU architecture with the DPU accelerator attached as a standard peripheral to the processor}
    \label{fig:datapath}
    \vspace{-15pt}
\end{figure}

Figure~\ref{fig:datapath} shows the proposed ECU architecture of the IDS-enabled ECU on a Xilinx Zynq Ultrascale+ device. 
The Zynq device integrate capable ARM processors connected to a host of hardened memory mapped peripheral logic and interface protocols within the processing system (PS) section of the device. 
Any custom peripheral can be integrated into the programmable logic (PL) region, and wired up the PS using high or low performance Advaned eXtensible Interface (AXI) ports and can be accessed as memory mapped devices from the software application on the processor.
As shown in figure~\ref{fig:datapath}, we are using an off-the-shelf CAN controller IP in the PL which implements the CAN and CAN-FD variants~\cite{xilinxcanfd}. 
Though the hardened CAN controller block in the PS could be used in this experiment, the PL integration provides a pathway for updating to future network standards.
The PL also instantiates a Xilinx DPU~\cite{xilinxdpu} block as our dedicated ML accelerator which can be configured and controlled using the Vitis AI Runtime (VART) APIs from the PS. 
The standard Vitis-AI flow is used to automatically generate the DPU IP along with wrappers, interconnects and the runtime elements from our high-level ML model described in TensorFlow (see Sec.~\ref{subsec:MLmodel}).
The model fetches CAN message information from the PS over the AXI interface when ready and provides the data back to the software using an interrupt-based read back, allowing the invoking task to run in a non-blocking fashion. 
Figure~\ref{fig:datapath} also shows the packet flow in the receive direction within the ECU. 
When the CAN interface receives a valid CAN packet, the software task on the PS reads the message like in every ECU application to process the information and take appropriate actions. 
Additionally, the DPU task sends the packet information to the IDS accelerator to detect any malicious nature that could be embedded in the message sequence. 
To achieve this, the software task managing the IDS allocates a \textit{n} 16-bit (or 32-bit in case of extended frames) FIFO-style buffer which is used as the input memory location for the DPU. 
At the reception of a new packet, the software task extracts the ID bits, updates the buffer and offloads the computation to the DPU using the non-blocking \emph{execute} command; the DPU runs the model on the packet information and interrupts the PS with a completion status.  
This scheme allows for a seamless integration with the normal event/time-triggered tasks to be executed by the ECU functions.

\subsection{CNN-based IDS}
\label{subsec:MLmodel}
To determine the best ML model, we explored different network architectures with varying complexity to find a baseline model that offers high inference accuracy at minimal complexity. 
We used our sequences of \textit{n} successive CAN IDs as the input feature from which the model extracted optimal features to learn.
Our model exploration was also guided by the network layers supported by the Vitis-AI framework. 
Based on our exploration, we chose a deep CNN (dCNN) network as it provided the best combination of accuracy and computational efficiency across multiple attack vectors.
The model consists of 8 \emph{Conv2D} layers implemented with 40, 80, 120, 160, 200, 240, 256 \& 512 filters at each layer, each filter having a dimension of 3$\times$3.
The time-series data composed of \textit{n} sequential CAN IDs is fed as input to the first convolutional layer with 40 filters. 
Subsequent layer(s) operate on the output of the previous layer, with increasing complexity to extract detailed features.
Batch Normalisation and Dropout layers were used between the convolutional layers to prevent over-fitting and to improve the learning efficiency during the training phase. 
The output layer uses a \emph{softmax} function at the output of the final dense layer to estimate the probability of an infected message. 
The model is defined in TensorFlow using standard TF functions and nodes.

\subsection{Dataset and Training}
\label{subsec:dataset}
We use the open Car Hacking dataset for intrusion detection for training our model and to test its performance~\cite{song2020vehicle}. 
The dataset provides a labelled set of normal and attack messages which were captured via the Onboard Diagnostic (OBD) port in an actual vehicle, with attack messages injected in real-time. 
The dataset includes DoS, Fuzzing and Spoofing message injections allowing us to validate the detection accuracy across these different attacks. 
An extract from the dataset is shown in table~\ref{table:4}, showing the CAN ID field, control field and the actual data segment.
We split the dataset as 80:15:5 for training, validation, and testing respectively, allocating the large section to training and optimisation of the quantised network. 

\begin{table}[t!]
\centering
\caption{An extract from the open Car hacking dataset which is used for our testing and evaluation}
\scalebox{0.92}{
\begin{tabular}{@{}lrrr@{}}
\toprule
\textbf{Time}     & \textbf{ID} & \textbf{DLC} & \textbf{Data}           \\ \midrule
\textbf{$\ldots$} \\
1478198376.389427 & 0316            & 8            & 05,21,68,09,21,21,00,6f \\
1478198376.389636 & 018f            & 8            & fe,5b,00,00,00,3c,00,00 \\
1478198376.389864 & 0260            & 8            & 19,21,22,30,08,8e,6d,3a \\ 
\textbf{$\ldots$} \\
\bottomrule
\end{tabular}}
\vspace{-0.45cm}
\label{table:4}
\end{table}

We pre-process the dataset prior to training to mimic the dataflow the model will obtain as its input, when integrated into the ECU. 
To achieve this, each message is converted to the equivalent binary value (modeling reception from the CAN interface) from which the ID's are extracted. 
To generate our FIFO-style operation, groups of \textit{n} sequential IDs are stacked to form the input tensor for the IDS. 
We set the value of \textit{n} as 4 for the DoS and Fuzzing attacks, 8 for Spoofing attacks for optimal performance based on our observation from the design space exploration step.
Each layer of this stack forms the combined input shape for the first layer, which is reshaped to feed into the exact channels for training and testing.

To train the model, we used adam optimizer with binary cross-entropy loss function. 
The learning rate was set to 0.0001 to allow for slower learning which aids in reducing loss of accuracy when quantising the pre-trained model~\cite{wu2018training}.
The training was performed on a workstation class machine with an Intel i9-9820X and an Nvidia A6000 GPU.
The trained model was exported as an \textit{h5} checkpoint for optimisation by the Vitis-AI flow, which transforms the model as a series of executable instructions on the DPU IP.
The models are quantised to 8-bit precision and optimised post quantisation for deployment by the Vitis-AI flow, also generating the APIs for moving data from and to the DPU from the PS.

\section{Experiments}\label{sec:experiments}

For our experiments, we use the testing split of the dataset which is similarly preprocessed to convert into the binary format. 
We mimic each packet arrival by extracting each message from this dataset, populating the FIFO-style buffer with the CAN ID from the current message and then invoking the DPU to process the contents in the buffer. 
We use the Python APIs generated by the Vitis flow for our testing, running on top of a Linux image on our XCZU7EV device on the ZCU104 development board.
To model other platform choices that could be explored for IDS, we also test the full-precision model (pre-quantisation) on an Intel Core i9-10900 CPU (10 cores) and on an Nvidia RTX A6000 GPU. 
The PyTorch model (pre-quantisation) is evaluated on the A53 ARM cores of the PS, modeling a typical dedicated software-driven IDS implementation on an automotive-grade processor.
In each case, we extract each message from the dataset and compose the time-series binary CAN ID sequence, which is provided as the input to the ML model. 
The FPGA accelerator (DPU) uses a 300\,MHz interface clock and a 600\,MHz DSP core clock, while the ARM PS is clocked at 1.2\,GHz. 
The Core i9 CPU has a base clock of 2.8 GHz and the Nvidia GPU has a base operating frequency of 1410\,MHz. 
We quantify the accuracy of inference by evaluating precision, recall, F1 rates as well as the false positive and false negative rates (FPR and FNR respectively) for our pre-quantised and quantised variants (on GPU and FPGA respectively).
We quantify the processing latency and power consumption for a single execution of a model on all platforms to quantify the inference time required for each new CAN message.
We configure a B4096 DPU with a single compute core and single execution thread to determine our baseline performance which is used to compare against competing approaches on alternate platforms and in literature. 
Further, we evaluate the scalability of the DPU-based IDS by evaluating the inference latency, resource overheads, and power consumption among different DPU configurations. 
All measurements are averaged over multiple runs, each using 50000 messages from our test split of the dataset. 

\subsection{Accuracy}
To quantify the possible loss in accuracy due to quantisation, we computed the precision, recall, and F1 rates for our quantised implementation on the FPGA-accelerator (QdCNN in the table) and compare it to our golden model (pre-quantisation TF model) on the GPU. 
The results, compiled in table~\ref{table:preqcomp}, show that the quantised model has nearly identical inference accuracy across different attack vectors, with notable improvement on the Fuzzy attack. 

\begin{table}[t!]
\centering
\caption{Inference accuracy metrics of the our model pre and post quantization on the four datasets} 
\scalebox{0.90}{
\begin{tabular}{@{}lllllll@{}}
\toprule
\textbf{Attack} & \textbf{Model}  & \textbf{Precision} & \textbf{Recall} & \textbf{F1} & \textbf{FPR(\%)} & \textbf{FNR(\%)} \\
\midrule
\multirow{2}{*}{DoS} & Pre-Q          & 0.9997             & 1               & 0.9998    & 0.01 & 0  \\
& QdCNN         & 0.9997             & 1               & 0.9998 & 0.01 & 0      \\
\multirow{2}{*}{Fuzzy} & Pre-Q          & 0.9982             & 0.7635          & 0.8652    & 0.04 & 23.65  \\
& QdCNN         & 0.9938             & 0.9851          & 0.9894   & 0.18 & 1.49   \\
\multirow{2}{*}{RPM-Spoof} & Pre-Q          & 0.9714             & 0.9872          & 0.9793 & 1.05 & 1.28     \\
& QdCNN         & 0.9764             & 0.9841          & 0.9803  & 0.86 & 1.59    \\
\multirow{2}{*}{Gear-Spoof} & Pre-Q          &  0.9755             &  0.9922          &  0.9838 & 1.58 & 0.78     \\
& QdCNN         &  0.9762             &  0.9940          & 0.9850 & 1.54 & 0.6    \\
\bottomrule
\end{tabular}}
\vspace{-0.1cm}
\label{table:preqcomp}
\end{table}
 
Table~\ref{table:dcnncomp} compares the accuracy metrics of our QdCNN model on the DPU against state-of-the-art works discussed in literature:  GIDS~\cite{seo2018gids}, DCNN~\cite{song2020vehicle} \& iForest~\cite{de2021efficient}.
As observed, the DCNN model has the highest accuracy numbers among the competing schemes.
Our quantised hardware accelerator offers nearly identical detection performance as the DCNN model in the DoS ($\pm$ 0.03\%) and Fuzzing attacks ($<$ 0.6\%) while falling slightly short ($\approx$ 2.2\%) in spoofing attacks.
Across the other models, we observe that our QdCNN is on par or better across all metrics for DoS and Fuzzing  attacks, with 1.4\% performance dip in the spoofing attack. 

\begin{table}[t!]
\centering
\caption{Accuracy metric comparison of our QdCNN FPGA accelerator to other state-of-the-art methods}
\scalebox{0.9}{
\begin{tabular}{@{}llllll@{}}
\toprule
\textbf{Attack}  & \textbf{Model} & \textbf{Precision} & \textbf{Recall} & \textbf{F1}  & \textbf{FNR} \\
\midrule
\multirow{4}{*}{DoS} & GIDS                  & 0.968                & 0.996          & 0.981  &   -   \\ 
& DCNN                  & 1.0                & 0.9989          & 0.9995  & 0.13\%     \\
& iForest                  & -                &   -       &  - &  -    \\ 
& \textbf{QdCNN (DPU)}                  & 0.9997             & 1               & 0.9998  & 0\%     \\
\midrule
\multirow{4}{*}{Fuzzy} & GIDS                  & 0.973                & 0.995          & 0.983   & -    \\ 
& DCNN                 & 0.9995             & 0.9965          & 0.9980  & 0.5\%     \\
& iForest                  & 0.9507                & 0.9993          & 0.9744  &    -  \\
& \textbf{QdCNN (DPU)}            & 0.9938             & 0.9851          & 0.9894   & 1.49\%     \\
\midrule
\multirow{4}{*}{RPM-Spoof} & GIDS                  & 0.983                & 0.99          & 0.986  &    - \\
& DCNN                 & 0.9999             & 0.9994          & 0.9996  & 0.11\%     \\
& iForest                  & 0.9897                & 1          & 0.9948  &    -  \\
& \textbf{QdCNN (DPU)}          & 0.9764             & 0.9841          & 0.9803    & 1.59\%    \\
\midrule
\multirow{4}{*}{Gear-Spoof} & GIDS                  & 0.981                & 0.965          & 0.972  &  -   \\
& DCNN                 & 0.9999             & 0.9989          & 0.9994  & 0.12\%     \\
& iForest                  & 0.9479                & 1          & 0.9733  &  -     \\
& \textbf{QdCNN (DPU)}              & 0.9760             & 0.9940          & 0.9850  & 0.6\% \\
\bottomrule
\end{tabular}}
\label{table:dcnncomp}
\end{table}

Table~\ref{table:confmatrix} shows the confusion matrix resulting from our evaluation of our QdCNN model on the Zynq ECU. 
The confusion matrix captures the performance of the quantised model beyond raw accuracy numbers of the model. 
As we inferred from our comparison, the model performs quite well in DoS and Fuzzing attacks, while we do observe higher classification errors in case of spoofing attacks. 
We believe that this performance could be further improved through focused fine-tuning within the Vitis-AI post-quantisation flow. 

\begin{table}[t!]
\centering
\caption{Confusion matrix capturing the classification results of our QdCNN on the DoS, Fuzzy, and Spoofing attacks}
    \scalebox{1}{
        \begin{tabular}{@{}llrr@{}}
            \toprule
            \multirow{2}{*}{\textbf{Attack}} & \multirow{2}{*}{\textbf{Message Type}}  & Predicted  & Predicted  \\ 
            & & Attack & Normal \\
            \midrule
          \multirow{2}{*}{DoS} & True Attack    & 16269                                & 0                                   \\ 
            & True Normal    & 5                                  & 33726                                \\ 
            \multirow{2}{*}{Fuzzy} & True Attack    &    11012                            &   166                                 \\
           &  True Normal    &    69                              &  38753                              \\ 
            \multirow{2}{*}{RPM-Spoof} & True Attack    & 13080                                 & 211                                  \\
            & True Normal    & 316                                  & 36393                                \\ 
            \multirow{2}{*}{Gear-Spoof} & True Attack    &   19299                                & 117                                  \\
            & True Normal    & 471                                 &  30113                              \\ 
            \bottomrule
        \end{tabular}}
        \vspace{-0.1cm}
\label{table:confmatrix}
\end{table}

\subsection{Inference Latency, Power \& Resource consumption}
To show the potential of the tightly coupled IDS, we quantify the per-message inference latency of the QdCNN model on our proposed ECU architecture and compare them against native precision models on other platforms: ARM CPU, desktop-class Intel i9-10900 CPU and an Nvidia RTX A6000 GPU. 
We also compare the observations against competing GPU/Rasberry Pi implementations reported in literature. 
In each case, we measure the time taken by the inference model by measuring the start and end times just before and after the execution calls are made. 
We measure the latency using back to back invocations of the ML model (as soon as the current execution is completed) and average them over 50000 runs.
The results can be observed in table~\ref{table:lat_power}.
The ARM CPU incurs~22\,ms for inference on average making it a poor candidate for line-rate IDS in CAN networks. 
While the GPU and Intel i9 CPU offers improved latency, the tightly coupled DPU accelerator in the proposed ECU architecture offers over 2$\times$ improvement compared to the GPU in case of per-message execution.
It should be noted that the PCIe transactions involved in GPU are very inefficient for short data transfers, which impacts the latency in case of the GPU model. 
Switching to batch mode on the GPU (executing all 50000 messages at once), it is possible to reduce the average execution to 0.195\,ms; however, it requires a significant number of CAN messages to be accumulated (batch size) before any potential threat can be identified, limiting its application as a line rate IDS. 

\begin{table}[t!]
\centering
\caption{Per-message latency and measured power on different platforms}
\begin{tabular}{@{}lrrr@{}}
\toprule
\multirow{2}{*}{\textbf{Platform}} & \textbf{Latency} & \multicolumn{2}{c}{\textbf{Measured Power}} \\ \cmidrule{3-4}
    & (ms) & Idle (W) & Active (W) \\ \midrule
Zynq-PS &  22.32   &  2.58      & 2.96 \\
i9-10900  & 1.8   & 35   &  152 \\
RTX A6000  & 1.1  & 22  &  96 \\
\textbf{DPU Accelerator (Zynq-PL)} & 0.53  & 4.37  &  5.76 \\ 
\bottomrule
\end{tabular}
\label{table:lat_power}
\end{table}

We also quantify the active power consumption of each platform while executing the inference model.
Here, we monitor the idle power draw on each platform with standard background tasks active and then measure the active power consumption when the model is invoked to arrive at the power consumed by the platform while executing the model. 
In case of the GPU, we used the Nvidia plugin (nvtop) to measure the GPU power consumption directly allowing it to be isolated from the host machine's power consumption. 
On the Zynq device, we use the on-device power rail monitors to measure accurate device level power consumption through the PMBus APIs. 
Table~\ref{table:lat_power} tabulates the power consumption measured across the different target architectures. 
We observe that the idle power consumption on the Zynq ZCU104 development board with the PS running Petalinux is 2.58\,W, while loading the PL region with our DPU bitstream raises the idle power consumption to 4.37\,W.
The ARM core on the PS consumes 2.96\,W while executing the model, while offloading the execution to the DPU results in a total power consumption of 5.76\,W by the Zynq device (and peripherals).
The proposed DPU offload in our ECU architecture can thus reduce the power consumption by 94\% 
while achieving lower per-message latency and nearly identical detection performance as a dedicated GPU-based IDS.
Note that the offload causes negligible overhead in the ARM cores (as opposed to software IDS deployment, consuming an estimated 4.5\,W during single core operation on a Raspberry-Pi 4 at full utilisation), allowing IDS capability to be integrated with any critical ECU function on the Zynq device with minimal effort. 


\begin{table}[t!]
\centering
\caption{Per-message latency comparison against state of the art IDSs (GPU/Raspberry Pi) reported in literature}
\scalebox{0.95}{
\begin{tabular}{@{}ll@{}}
\toprule
     & \textbf{Latency (ms)}  \\ \midrule
MLIDS~\cite{desta2020mlids} & 275 (per CAN message, GPU)\\
GIDS~\cite{seo2018gids} & 5.89 (block of 64 CAN frames, GPU) \\ 
DCNN~\cite{song2020vehicle} &  5 (block of 29 CAN frames, GPU)      \\
MTH-IDS~\cite{yang2021mth} &  0.574 (per CAN message, Raspberry Pi)      \\
\textbf{QdCNN on DPU} & 0.53 (per CAN message, DPU on Zynq FPGA)    \\ 
\bottomrule
\end{tabular}}
\vspace{-0.35cm}
\label{table:latcomp}
\end{table}

We also compare the latency of the proposed IDS-ECU architecture against other implementation platforms discussed in literature~\cite{desta2020mlids,seo2018gids,song2020vehicle,yang2021mth} in Table~\ref{table:latcomp}.
The DCNN \& GIDS approach operates on a block of 29 \& 64 CAN messages at each execution, consuming 5\,ms \& 5.89\,ms to process the block on an Nvidia Tesla K80 GPU \& GTX 1080 GPUs respectively.
On average, this leads to a processing latency of 0.17\,ms \& 0.09\,ms per message, at the expense of higher waiting time for arrival of the batch of messages. 
In~\cite{desta2020mlids}, the authors measure the per-message latency of 275\,ms  for their model, when executed on an Nvidia Titan X (GTX 200) GPU. 
In~\cite{yang2021mth}, the authors report a per-message latency of 0.574\,ms when their lightweight multi-model stacked IDS is executed on a Raspberry Pi 3 device.
In comparison, our integrated IDS-ECU incurs 0.53\,ms per-message latency, and in combination with the high detection accuracy across multiple attack modes, lower power consumption (vs GPU) and the ability to tightly couple IDS with ECU function, lends itself as the ideal architecture for real-time IDS on CAN networks. 

We also quantify the resource consumption of the custom modules on the programmable logic part of our proposed IDS-ECU on the Zynq XCZU7EV device. 
The resource consumption and the overall utilisation of the device is shown in table~\ref{table:resourceutilization}. 
It can be seen that the largest DPU model (B4096) that can be deployed on this device consumes most of the resources as expected; however, it still leaves enough resources for other custom accelerators which could be integrated to the ECU to offload parts of its tasks.


\begin{table}[t!]
\centering
\caption{Resource utilisation breakdown for the PL modules of our proposed CAN IDS-ECU (XCZU7EV)}
\scalebox{0.95}{
\begin{tabular}{@{}lrrrr@{}}
\toprule
\textbf{Node} & \textbf{LUT} & \textbf{FF} & \textbf{BRAM/URAM} & \textbf{DSP} \\
\midrule
DPU (B4096)  & 48313         &  97508          & 84.5/ 46           &  690 \\
CAN FD        & 1932         & 2110        & 6/0             & 0              \\
\midrule
Overall & 64660  & 117562  & 116.5/46  & 704 \\
(\% usage) & (28.06\%) & (25.51\%) & (37.34\%/47.92\%) & (40.74\%) \\
\bottomrule
\end{tabular}}
\label{table:resourceutilization}
\end{table}

\subsection{Performance scaling using alternate DPU configurations}
Vitis-AI offers a range of DPU configurations allowing the designer to trade-off performance for resource and power consumption.  
The different DPU configurations tune the amount of processing and input-output parallelism, incurring higher resources and power consumption for higher performance. 
Table~\ref{table:dpucomparison} shows various performance/utilisation/power metrics when our QdCNN model is run on different DPU configurations on the Zynq XCZU7EV device.
%
The increased performance of B4096 DPU improves the latency per message by 34.65\% while utilising more DSP and distributed memory (BRAM/URAM) blocks and increasing the power consumption from 3.84\,W to 5.76\,W compared to the baseline B512 DPU.
From table~\ref{table:dpucomparison},  B1152 DPU offers the best trade-off, achieving nearly 22\% reduction in message latency at under 6\% increase in DSP utilisation and around 15\% increase in power consumption compared to the B512 DPU. 

\begin{table}[t!]
\centering
\caption{Performance scaling of the QdCNN model using different DPU configurations: impact on per message latency (ms), resource (\%), \& power consumption (W)}
\scalebox{0.8}{
\begin{tabular}{@{}lcccccccc@{}}
\toprule
\multirow{2}{*}{\textbf{DPU}} & \multirow{2}{*}{\textbf{Latency}} & \multicolumn{5}{c}{\textbf{\% Resource Utilisation}} & \multicolumn{2}{c}{\textbf{Power (W)}} \\ \cmidrule{3-9}
 & (ms) & LUT & FF & DSP & BRAM & URAM & Idle & Active \\ \midrule
B512           & 0.81                  &  12.46 & 7.96 &6.37 &13.94 &12.5  &  2.87 & 3.84 \\
B800           & 0.70                &  13.55&9.47&9.09&15.54&29.17   &  3.14 & 4.15  \\
B1024           & 0.81               &  15.55&11.03&12.62&26.76&14.58  &  3.17 & 4.29   \\
B1152           & 0.63                &  14.67&10.83&12.27&18.91&33.33   &  3.16 & 4.42  \\
B1600           & 0.70                 &  17.14&13.57&18.06&29.97&33.33   &  3.40 & 4.83  \\
B2304           & 0.63                &  18.82&15.71&24.42&36.54&37.50   &  3.68 & 4.98  \\
B3136           & 0.58                &  20.84&18.16&31.71&42.95&41.67   &  3.91 & 5.46  \\
B4096           & 0.53                &  23.25&22.22&39.93&47.12&47.92   &  4.37 & 5.76  \\
\bottomrule
\end{tabular}}
\vspace{-0.4cm}
\label{table:dpucomparison}
\end{table}

\section{Conclusion}\label{sec:conclusion}
In this paper, we explored a hybrid-FPGA-based ECU architecture which uses a machine learning model to detect onset of intrusions on a CAN network. 
We utilise Xilinx's Vitis-AI flow to quantise, optimise and compile a deep-CNN model defined in TensorFlow for execution on an off-the-shelf DPU accelerator IP attached to the ECU.   
Our evaluation shows that the tightly coupled accelerator model offers an average accuracy of 99.32\% across multiple attack modes, almost identical to state of the art ML-based IDS models (on GPU) described in the literature.
Furthermore, the proposed architecture offers nearly 51\% reduction in average latency per message over the GPU deployment of our model and comparable or better than competing models in literature, while consuming less than 6\% of the power consumed by the GPU deployments.
We believe that the proposed architecture can pave way towards further research on line-rate distributed intrusion detection for CAN and Automotive Ethernet.

\section{Acknowledgement}
This research was supported by grants from NVIDIA and utilised NVIDIA RTX A6000 GPU.
\bibliographystyle{IEEEtran}
\bibliography{references.bib}

\end{document}